\documentclass[twocolumn,prl,showpacs,amsmath,amssymb,nofootinbib,superscriptaddress,preprintnumbers]{revtex4}

\usepackage{dcolumn}
\usepackage{amsmath}
\usepackage{rotating}
\usepackage{amsfonts}
\usepackage{amssymb}
\usepackage{subfigure}

\usepackage{graphicx}
\usepackage{psfig,color}

\newcommand{\be}{\begin{equation}}
\newcommand{\ee}{\end{equation}}
\def\bea{\begin{eqnarray}}
\def\eea{\end{eqnarray}}

\begin{document}

\preprint{IPPP/07/22}
\preprint{DCPT/07/44}

\title {Scattering amplitudes in strongly coupled N=4 SYM\\ 
from semiclassical strings in AdS }

\author{Steven Abel, Stefan F{\"o}rste, Valentin V. Khoze}

\medskip

\affiliation{Institute for Particle Physics Phenomenology,
Durham University, Durham, DH1 3LE, UK}

\medskip

\begin{abstract}
Very recently in \cite{AM} Alday and Maldacena gave a string theory prescription for computing (all) planar amplitudes
in ${\cal N}=4 $ supersymmetric gauge theory at strong coupling using the AdS/CFT correspondence. These amplitudes
are determined by a classical string solution and contain a universal exponential factor involving the action of the 
classical string. 
On the gauge theory side, expressions for perturbative amplitudes at strong coupling were previously proposed only for specific
helicities of external particles -- the maximally helicity violating or MHV amplitudes.
These follow from the exponential ansatz of Bern, Dixon and Smirnov \cite{BDS} for MHV amplitudes in ${\cal N}=4 $ SYM.
In this paper we examine the amplitudes dependence on helicities and particle-types of external states.
We consider the prefactor of string amplitudes and give arguments suggesting that the prefactor at strong coupling should be 
the same as the Yang-Mills tree-level amplitude for the same process. This implies that scattering amplitudes in ${\cal N}=4 $ SYM
simplify dramatically in the strong coupling limit. It follows from our proposal that in this limit
all (MHV and non-MHV) $n$-point amplitudes are given by the (known)
tree-level Yang-Mills result times the helicity-independent (and particle-type-independent) universal exponential.

\end{abstract}


\maketitle

In Ref.~\cite{AM} Alday and Maldacena uncovered how the AdS/CFT string -- gauge theory duality can be used 
to determine all $n$-point scattering amplitudes in ${\cal N}=4 $ supersymmetric gauge theory at strong coupling.

Until now it was unclear if
the AdS/CFT correspondence can address scattering amplitudes in gauge theory directly.
These amplitudes should correspond to scattering of open strings with ends living on $N$ D3 branes.
However in the Maldacena $\alpha' \to 0$ decoupling limit \cite{Maldacena} the D3 branes essentially disappear giving 
rise to the $AdS_5 \times S^5$ geometry which serves as the target space for type IIB theory of (closed) strings.
The well-known holographic relation \cite{Witten,GKP} relates Green functions of chiral primary composite operators 
in SYM to interactions of type IIB supergravity states in $AdS_5 \times S^5$, but does not address the on-shell
scattering amplitudes in SYM.

In the Alday-Maldacena approach \cite{AM}, as will be reviewed below, 
the open strings which correspond to gluons (and their superpartners)
end on the infrared D3 brane. This infrared brane is placed in the $AdS_5$ space at a large fixed value
$Z_{IR}$ of the radial coordinate, and extends along the four worldvolume directions $X^\mu$. The infrared
brane plays the role of the infrared cutoff in gauge theory. Scattering amplitudes of massless on-shell states are
infrared divergent in gauge theory and cannot be defined without an infrared cutoff. Infrared-regularised
amplitudes in gauge theory are used at intermediate stages to calculate infrared-safe physical observables,
such as jet cross sections, etc. In \cite{AM} scattering amplitudes of open strings in $AdS_5 \times S^5$
which end on the infrared D3 brane are identified with the IR-regularised amplitudes in the ${\cal N}=4 $ SYM.
Taking $Z_{IR} \to \infty$ removes the IR cutoff and renders these amplitude IR divergent.

At the leading order in strong coupling, $\lambda \to \infty$, scattering amplitudes $A_n$
are dominated by a single
classical string configuration whose boundary conditions are determined by the external momenta $p_1,\ldots,p_n$
as explained in \cite{AM}. The colour-ordered planar scattering amplitudes of $n$ gluons with momenta $p_i$
and helicities $h_i=\pm$ at strong coupling are of the form \cite{AM},
\be
A_n(p_1,h_1,\ldots,p_n,h_n) = K \,e^{i \sqrt{\lambda} S_{cl}} = K \,e^{-\frac{\sqrt{\lambda}}{2\pi}{\rm Area}_{cl}}\ ,
\label{expS}
\ee
where $S_{cl}$ is the worldsheet action evaluated on the classical solution. It is given by the
area of the minimal surface ${\rm Area}(p_1,\ldots,p_n)_{cl}$
in $AdS_5$ that ends on the boundary of the string worldsheet, with the prescribed boundary
conditions determined by the external momenta $p_i$. 
The exponent in \eqref{expS} is universal: 
every $n$-point amplitude (for fixed $n$) contains the same function $e^{-\frac{\sqrt{\lambda}}{2\pi}{\rm Area}_{cl}}$
of the $n$ external momenta. 
The prefactor $K$ in \eqref{expS} is non-universal: it depends 
on the helicities 
(and particle types) of the external states in $A_n$ as well as on the kinematics. Thus, $K$ distinguishes between
specific $n$-point amplitudes, and has to be determined for each amplitude in order
to, for example, calculate cross sections. 
In the $\lambda \to \infty$ limit 
the entire $\lambda$ dependence of the amplitudes is contained in $e^{-\frac{\sqrt{\lambda}}{2\pi}{\rm Area}_{cl}}$
while the prefactor $K$ is $\lambda$-independent.

The authors of \cite{AM} have concentrated on the universal exponent in \eqref{expS}. They have computed it explicitly
for 4-point amplitudes, and also have studied its infrared properties for general $n$-point amplitudes.
The ${\rm Area}_{cl}$ is infinite when $Z_{IR}\to \infty$ in agreement with the fact that the amplitudes are IR
divergent. When the IR regulator is present, it was shown in \cite{AM} (both using the explicit cutoff and the dimensional
reduction schemes) 
that the infrared properties of $e^{-\frac{\sqrt{\lambda}}{2\pi}{\rm Area}_{cl}}$ are in precise agreement
with the expected IR behaviour of resummed perturbative amplitudes in gauge theory \cite{MS,STY,C,BDS}.

For 4-point amplitudes the classical string action in the exponent of  \eqref{expS} 
was computed explicitly in \cite{AM} using dimensional reduction to $D=4-2\epsilon$ dimensions to order 
$\frac{1}{\epsilon^2}+\frac{1}{\epsilon}+\epsilon^0$. Agreement was found with  
the strong coupling limit of the previously known gauge theory result for $A_4$ due to 
Bern, Dixon and Smirnov (BDS) \cite{BDS}.
This fact, together with the matching of the IR behaviour for all $n$-point amplitudes provides a 
non-trivial test of the AdS/CFT result \eqref{expS} and, perhaps more importantly, explains from the string theory
perspective the {\it exponentiated} form of SYM amplitudes. In perturbative gauge theory this exponentiation 
is a consequence of the
up to now mysterious iterative structure of MHV loop amplitudes 
\cite{ABDK},\cite{BDS,Cachazo:2006mq,Cachazo:2006tj,Bern:2006vw,Bern:2007ct}
in planar perturbative ${\cal N}=4$ SYM.

The purpose of this letter is to determine the prefactor $K$ of the $n$-point amplitudes in \eqref{expS}.
As already mentioned, $K$ is the factor which distinguishes the amplitudes with different helicities (and 
particle states) and is required  
for any cross section-type calculation. We will  argue that for all $n$-point amplitudes $K$ is given by 
the corresponding tree-level results, $K \propto A_n^{\rm tree}$ which are either known or can be easily computed
using the tree-level MHV rules \cite{CSW,Khoze,GGK} in perturbative ${\cal N}=4$ SYM. 
More precisely, $K = A_n^{\rm tree} e^{-S_0}$, where $S_0$ does not depend on $\lambda$ and is also independent 
of the helicities of external particles.
In string theory
$S_0$ arises from the subleading corrections
to $\sqrt{\lambda} S_{cl}$ in the effective action.
If the BDS conjecture holds for all $n$-point MHV amplitudes in gauge theory,
$S_0$ can be easily determined by comparing to those.
In any case, if our approach is valid, it represents
arbitrary (MHV or non-MHV) gauge theory $n$-point planar amplitude 
at strong coupling in the factorised form
\be
A_n = A_n^{~ \rm tree} \,e^{i \sqrt{\lambda} S_{cl} -S_{0}} ,
\label{result}
\ee
where only the first factor on the RHS depends on helicities of external states.
This form of the answer is in agreement with what was known (or conjectured) previously
for MHV amplitudes in ${\cal N}=4$ SYM \cite{BDS} 
\be
A_n^{~ \rm MHV} = A_n^{~ \rm tree~ MHV} e^{F_n^{\rm BDS}(\lambda;p_i)} ,
\label{BDSans}
\ee
Equation \eqref{BDSans} is the exponential ansatz of Bern, Dixon and Smirnov,
conjectured to hold to all orders at weak coupling and continued to the strong coupling regime.
The fact that MHV loop amplitudes are proportional to the tree-level MHV amplitudes 
is a consequence of ${\cal N}=4$ supersymmetry (see e.g. Appendix E of Ref.~\cite{Bern:1998ug}
and can be understood from the fact that
$A_n^{~ \rm tree~ MHV}$ is given by a single term, see Eq.~\eqref{PTBG} below, which is
uniquely fixed by the kinematic limits and symmetries of ${\cal N}=4$ SYM.
It is known, however, that this factorisation does not hold order by order in $\lambda$ 
for non-MHV amplitudes, and \eqref{BDSans} is not valid beyond the MHV case at fixed values of
$\lambda$. However our result \eqref{result} implies that such factorisation does hold in 
the strong coupling limit $\lambda \to \infty$.

Our main goal is to find how string amplitudes depend on the helicities and types of 
external states. These states -- gluons and their ${\cal N}=4$ superpartners -- are massless
excitations of the open string. To proceed we first need to locate 
the worldsheet boundary of the open string and discuss where the vertex operators describing the external states
should be placed.
\begin{figure}
\begin{center}
{\includegraphics[width=8cm]{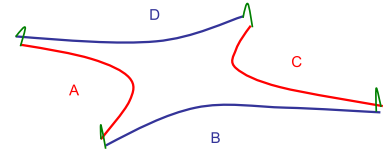}
\hspace{1cm}
\vspace{-16ex}
\caption[...]{Scattering of four open strings ending on $N$ coincident $D3$-branes. $A,B,C,D$ are the Chan-Paton indices
labelling the branes on which strings end. For future reference we choose external states described by open strings
with one end on the $N^{\rm th}$ brane, $B=D=N$ (shown in blue),
 and the other end on the remaining $N-1$ branes, $A,C=1,\ldots N-1$ (shown in red).
\label{fig:Fig-1-L}}} 
\end{center}
\end{figure}

The standard $AdS_5 \times S^5$ string theory description of conformal ${\cal N}=4$ SYM arises from considering a stack of $N$ coincident 
$D3$-branes in flat ten-dimensional IIB string theory and subsequently 
taking the large-$N$ near horizon limit \cite{Maldacena}.
Gluon scattering in gauge theory corresponds to a scattering of open strings with ends on the $D3$-branes from the stack,
as shown in Figure 1. External states are descried by the vertex operators 
$V (p) {T^{(a)}}^A_B$, where ${T^{(a)}}^A_B$ are the usual $SU(N)$ generators
which keep track of the Chan-Paton factors $A,B=1,\ldots,N$.

We will be following the philosophy of \cite{AM} where one $D3$-brane is separated from the stack of $N-1$
branes and placed at $Z=Z_{IR}$. This gives the Coulomb branch in gauge theory such that the states transforming
in the bifundamental\footnote{That is strings stretched between the $N-1$ stack and the single brane at $Z_{IR}.$}
 of $SU(N-1) \times U(1)$ become massive and all the remaining states of $SU(N)$ remain massless.
This procedure implements an IR regularisation for the amplitudes where all the {\it external} states are in the bifundamental
of $SU(N-1) \times U(1)$. In practical computations this prescription was actually not used in \cite{AM} (they instead adopted
a version of dimensional reduction), but for our task of keeping track of the external states,
we have to employ the 
more geometrical prescription in terms of the separated infrared probe brane. 

\begin{figure}
\begin{center}
{\includegraphics[width=8cm]{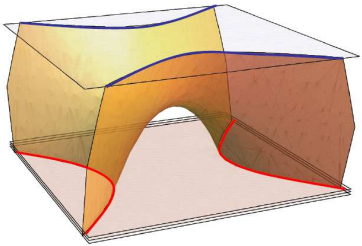}
\hspace{1cm}
\vspace{-6ex}
\caption[...]{Scattering of open strings stretched between the separated IR brane and the stack of $N-1$ $D3$-branes.
\label{fig:Fig-2-L}}} 
\end{center}
\end{figure}

In Figure 2 we show what happens to the amplitude in Figure 1 when the IR brane is separated from the $N-1$ stack.
The external states are those of the stretched strings.
We now take the Maldacena near-horizon
limit of the $N-1$ stack of $D3$-branes. The $N-1$ branes dissolve and generate the $AdS_5 \times S^5$ geometry with $N-1$ units
of flux through the $S^5$ as in the usual case, but the IR brane at $Z_{IR}$ remains and can be viewed as a probe brane.
Strings with both ends on the IR brane remain open strings, strings with both ends on the $N-1$ stack become IIB closed
strings in $AdS_5 \times S^5$. Our external states are strings which were 
stretched between the IR brane and the $N-1$ branes in Figure 2. 
These are replaced by the open--closed string interactions.
At tree level in the string coupling $g_{\rm st}$ these interactions are reflected by bending of the open string worldsheet 
in the classical $AdS_5$ background.
Thus, the string worldsheet of open strings previously stretched between the IR brane and the $N-1$ 
branes in Figure 2
is now bending into the bulk of the $AdS_5$ as shown in Figure 3. 
The vertex operators describing external states are located on the Dirichlet IR brane, which is the only brane remaining.

\begin{figure}
\begin{center}
{\includegraphics[width=8cm]{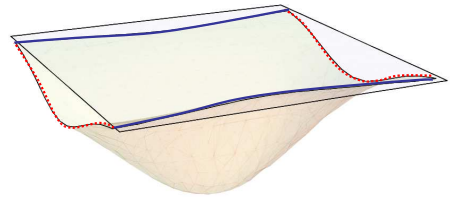}
\hspace{1cm}
\vspace{-16ex}
\caption[...]{In the Maldacena near-horizon
limit the $N-1$ stack dissolves into the $AdS_5 \times S^5$ geometry and the IR brane is the only brane remaining.
The stretched strings worldsheet in Figure 2 becomes
the open string worldsheet curved into the $AdS$ bulk. We show a slice of this worldsheet at finite values of $X^\mu$.
In the asymptotic region $X^\mu \to \infty$ the red dotted lines approach the IR brane so that all the asymptotic scattering states
are located on the IR brane.
\label{fig:Fig-3-L}}} 
\end{center}
\end{figure}

Normally one expects that the external states, being the states of the boundary conformal SYM theory, 
should live on on the boundary of the Anti de-Sitter $AdS_5$ space,
and this is where the boundary of the open string worldsheet must be.
In terms of Poincar\'e coordinates $(X^{\mu},Z)$ the $AdS_5$ boundary is spanned by coordinates $X^\mu$ and is placed
at the radial coordinate $Z \to 0$. 
However, our earlier discussion shows that the open string worldsheet in the Alday-Maldacena approach
ends on the IR brane at large values of $Z=Z_{IR}$ rather than at $Z \to 0$.
As explained in \cite{AM}, possible confusion is avoided
when one recalls that the 
$Z=Z_{IR} \to \infty$ surface also touches the boundary of $AdS_5$ at values of $X^\mu \to \infty$. Thus 
the boundary of $AdS_5$ is not only described by $Z \to 0$, but also by $Z=Z_{IR}$ at large values
of $X^\mu$. 

The reason why the correct description of gluon scattering is achieved in terms of open strings 
ending at $Z=Z_{IR}$ and not at $Z \to 0$ is most easily understood by comparing two problems:
one of gluon scattering in $SU(N)$
gauge theory, and the other of calculation of Wilson loops made by infinitely heavy quarks
in the fundamental representation of the $SU(N-1)$  SYM. Both problems are addressed by considering the worldvolume
actions of the string stretched between a single D3 brane put at a fixed value of $Z$ and the remaining 
$N-1$ D3 branes of the $SU(N)$ gauge theory. The mass of the stretched string goes to infinity when
$Z \to 0$ (giving an infinitely heavy fundamental quark state of $SU(N-1)$)
and this set-up corresponds to the Wilson loop calculation of \cite{Wilsonloop}. The alternative scenario
where the single brane is at $Z=Z_{IR}$ gives masses of stretched strings of order $m=1/Z_{IR}$
which correspond to light gluons (becoming massless when the IR regulator is removed, $1/Z_{IR} \to 0$).

Thus we have established that the asymptotic external states live on the boundary of $AdS_5$, which (up to a constant
rescaling by $Z_{IR}$) is the 4-dimensional Minkowski space.\footnote{Recall that the Dirichlet IR brane which defines the
string worldsheet boundary, approaches the boundary of $AdS_5$ at asymptoticly large $X$. Since the IR brane is placed at $Z_{IR}$,
the relevant to us component of the $AdS_5$ boundary is also at $Z_{IR}$. Hence the vertex operators \eqref{Voper} below
will also be at $Z_{IR}$.}
There we can use the standard flat space definition of 
vertex operators $V$. For a gluon state of momentum $p_i$ and helicity $h_i=\pm$ we use
\be
V(p_i) = \int d\tau \, e^{i p_i^\mu X_\mu (\tau)}\varepsilon_{\mu}^{\pm}(p_i) (\partial_{\tau} X^\mu (\tau) + \ldots) ,
\label{Voper}
\ee
where $\tau$ parameterises the boundary of the worldsheet $(\tau,\sigma)$,
 the polarisation vector of the gluon is $\varepsilon_{\mu}^{\pm}(p_i)$
and $X^\mu = X^\mu (\tau, \sigma=0)$ is taken at the boundary (with the radial coordinate $Z=Z_{IR}$).
The ellipses on the right hand side indicate the supersymmetric completion of the vertex operator.
These terms will not modify our conclusion for the prefactor.\footnote{It will turn out that the prefactor is 
determined in our approach by a corresponding tree-level open string amplitude in the $\alpha' \to 0$ limit
and in the flat background.
In this limit the amplitude is the same whether it is calculated in the non-supersymmetric open string theory,
or in the fully supersymmetrised version reflecting the fact that gauge theory tree-level amplitudes are not sensitive
to superpartners. (For fixed external states, superpartners can propagate only in loops and are decoupled at tree level.)}

As mentioned earlier,
the vertex operators \eqref{Voper} must be accompanied by the Chan-Paton factors of the external states of the form
${T^{(a)}}^A_B$ where the choice of our external states requires that one of the labels $A,B$ corresponds to the
IR brane, and the other runs from $1$ to $N-1$. 
The full Chan-Paton factors of the amplitudes in Figures 1, 2 remain unchanged after the Maldacena limit is taken
and are inherited by the amplitudes in Figure 3 even though the $N-1$ stack has disappeared.
The amplitudes can now be represented in the colour-ordered form 
and we can concentrate on the purely kinematic partial amplitudes, ignoring the colour $SU(N)$ structure.
Full amplitudes are obtained from the kinematic partial amplitudes in the standard way (see e.g. \cite{MP,LDTASI})
by multiplying them 
with known colour structures (traces of products of $T^{(a)}$'s) and summing over inequivalent permutations.
The general philosophy up to now was set up to address scattering amplitudes with very specific external states
-- those transforming in the bifundamental of $SU(N-1)\times U(1)$ -- which is a subset of the full set of states
in the adjoint representation of $SU(N)$. Then from the kinematic partial amplitudes arising in this approach
one can assemble the full $SU(N)$ amplitude for general $SU(N)$ external states by simply assuming the full
$SU(N)$ gauge invariance.
Moreover, we also expect that the original set-up itself can be extended to address more
general non-Abelian amplitudes. To this end one would have to pull out a few
(but still a fixed number) of distinct IR branes before taking the large-$N$ near horizon limit. 
The choice of external states dictates which IR branes should be selected for each process.
We will not pursue this any further at present.

The $n$-point open string partial amplitude is represented by the Polyakov's functional integral
\be 
A_n = \int DX \, V(p_1) \ldots V(p_n) \, e^{i \sqrt{\lambda} S[X]} ,
\label{Int}
\ee
where $\sqrt{\lambda} S[X]$ is the worldsheet action of the sigma model with the
$AdS_5 \times S^5$ target space.

The amplitude \eqref{Int} can be recast into the form
\begin{equation}
A_n = \prod_i \int d\tau_i \epsilon^{\pm}_\mu\left( p_i\right)
    \frac{\delta }{\delta J_\mu\left( 
    \tau_i\right)} e^{iW\left[ J\right]} ,
    \label{Ampl}
\end{equation}
with
\begin{eqnarray}
e^{iW\left[ J\right]} & = &
\int DX DZ \exp\left\{ i\sqrt{\lambda} S\left[ X,Z \right] +
   \sum_i i p^\mu_i X^\mu \left( \tau_i\right)\right. \nonumber \\ & & \left. +
    \int d\tau J_\mu \left( 
    \tau\right) \partial_\tau {X}_\mu \right\} . \label{eq:geni}
\end{eqnarray}
As in \cite{AM}, the $S^5$ sphere does not play an important role
for amplitudes of gluons, and 
the string solution is described by the $({X}^{\mu},Z)$ fields of $AdS_5$ 
where $\mu=0,\ldots,3$ and $Z$ is the radial coordinate. The bosonic 
action on the $AdS_5$ is
\be
S\left[ X,Z \right] = \int d\tau d\sigma \left(
\partial_\alpha X^\mu \frac{1}{Z^2}
\partial^\alpha X_\mu  + 
\partial_\alpha Z \frac{1}{Z^2} 
\partial^\alpha Z \right)
\label{SZX}
\ee

Now, in general performing the integrals in \eqref{eq:geni}
would be a difficult problem, because one would require an infrared-regulated solution
with a full dependence on the emission points $\tau_i$. However, we will argue that such a solution
can be obtained by patching a $\tau_i$-independent IR-regulated solution of \cite{AM} in the bulk
to a new $\tau_i$-dependent solution which we will construct near the boundary.

We start by revisiting the classical string solution
of \cite{AM}. This is an extremum of the exponent in \eqref{eq:geni} 
without sources, i.e. at $J=0$.
It is best described in terms of fields $(y^{\mu}, r)$ dual to $(X^{\mu}, Z)$ 
which are defined via \cite{AM}
\be
\partial_{\alpha} y^\mu = i \frac{1}{Z^2} \epsilon_{\alpha \beta} \partial_{\beta} X^\mu \ , \ \, 
r= \frac{1}{Z}
\ee
The boundary conditions that the original coordinate $X^{\mu}$ carries a momentum $p_i^\mu$ 
give the conditions
that $y^\mu$ jumps by an amount proportional to the momentum $p_i^\mu$ at the emission points. 
The simplest solution describing the 4-point amplitude in the special case of $s=t$ 
has the boundary conditions for $y^{\mu}$ depicted by four light-like segments in Figure 4 (a), and 
$r=r_{IR} \equiv 1/Z_{IR}$. Projection on the $(y_1,y_2)$ plane gives the square with the sides of length
$k=\sqrt{s/2}$ which by rescaling can be set to 2, as shown in Figure  4 (b). 

Ideally for our full action, one would like to instead
evaluate it in the conformal gauge 
retaining full dependence on $\tau_i$. One would place
each of the four vertex operators $V(\tau_i)$ on each side of the 
square.\footnote{For example at the insertion point
$\tau_1$ we have $y_1 =+1$ and $y_2$ changes from $-1$ to $+1$, dictated by $\Delta y_2(\tau_1) = \vec{k}_1 =2$.} 
More precisely, 
\be
y^\mu (\tau, \sigma=0) = \sum_i \theta(\tau-\tau_i) 
\frac{p_i^\mu}{\sqrt{\lambda}} \equiv \sum_i \theta(\tau-\tau_i) k_i^\mu
\ee
where $\theta$ is the step function and we have defined the rescaled momenta $k_i^\mu$. 
However the answer would be dominated by the classical action of \cite{AM}
which is independent of the $\tau_i$.

\begin{figure}
\begin{center}
\subfigure[]{
\includegraphics[width=4cm]{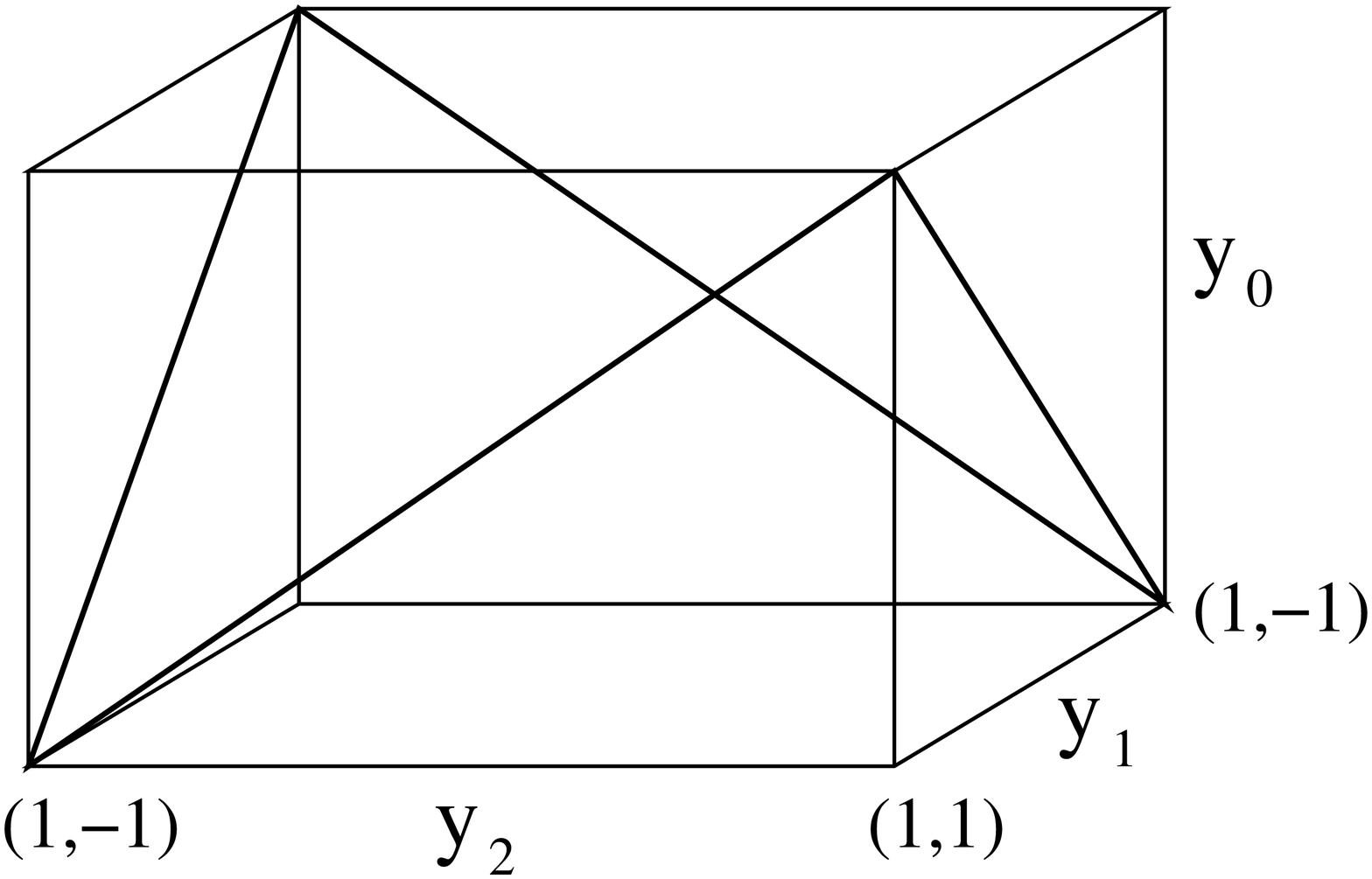}
\label{fig:QFT}}
\hspace{0.7cm}
\subfigure[]{
\scalebox{0.85}[0.85]{
\hspace{-0.9cm}
\includegraphics[width=4cm]{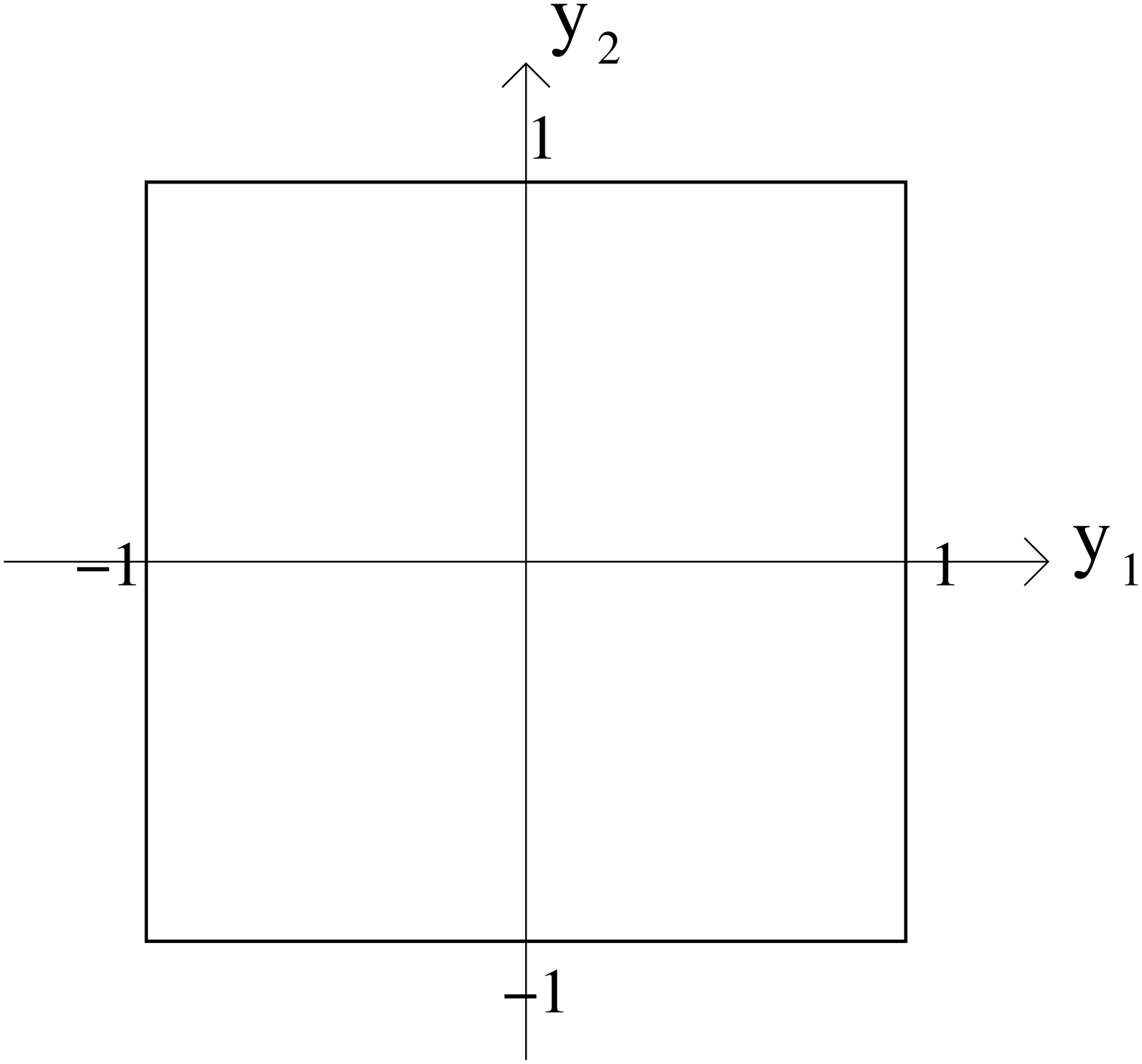}
}\label{fig:string}}
\end{center}
\vspace{-6ex}
\caption[...]{Four point scattering in the T-dual picture (with sides 
rescaled from $\sqrt{s}/4$ to unity.  
\label{fig:tdual}} 
\end{figure}
\begin{figure}
\begin{center}
{\includegraphics[width=5cm]{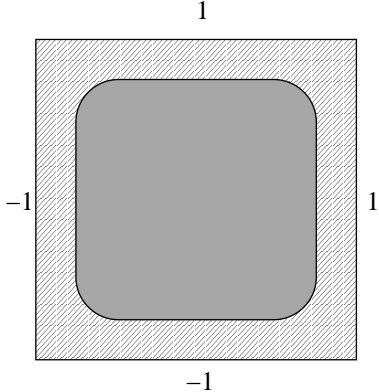}
\hspace{1cm}
\caption[...]{$(y_1,y_2)$-projection of the solution for four-point scattering showing the square boundary,
and the smoothed boundary. The shaded inner region 
(in grey) shows the IR regulated ``bulk'' contribution. The dotted outer 
region is where the patched solution is constructed
with $r \simeq r_{IR}$.
\label{fig:scatter}}} 
\end{center}
\end{figure}

Now consider the procedure for IR regulation.
The solution of \cite{AM} 
\be
y_0(y_1,y_2) = y_1 y_2 \ , \, \, 
r(y_1,y_2)= \sqrt{(1-y_1^2)(1-y_2^2)}
\label{solam}
\ee
does not have an explicit infrared cutoff; instead of boundary conditions
\bea
&&r(\pm 1,y_2)=r(y_1,\pm 1) =r_{IR} , \\
&&y_0(\pm 1,y_2)=\pm y_2 \ , \, \, y_0(y_1,\pm 1)=\pm y_2
\nonumber
\eea
the solution \eqref{solam} satisfies the boundary conditions with
$r_{IR}=0$.
In \cite{AM} the classical action on the solution \eqref{solam}
(and on more general boosted solutions with $s \neq t$)
was regulated by dimensional reduction with the IR divergences manifesting themselves as
$1/\varepsilon$ poles. Here however it is more illuminating to 
introduce an IR cutoff into \eqref{solam}
by limiting the range of integrations over $y_1$ and $y_2$ such that on the
boundary of the integration region, the function $r$ is constant $r_{IR} \neq 0$.
This defines the boundary in the form
\be
r_{IR}^2 = (1-y_1^2)(1-y_2^2) \ ,  \, \, 0< r_{IR} \ll 1
\label{bound}
\ee
which we plot in Figure 5. 
In terms of the explicit IR cutoff $r_{IR} =1/Z_{IR}$, the action 
on the solution \eqref{solam},\eqref{bound} is equivalent
to the double integral performed inside the boundary curve \eqref{bound} (the shaded region
in Figure 5.) 
In other words, by construction, the action in the inner region corresponds to an IR-regulated 
Alday-Maldacena universal exponent, albeit in the $Z_{IR}$ regularisation rather than the dimensional reduction  
scheme used in [1].

This regulation \eqref{bound} clearly excises an important region from the integration, 
namely the cusps of the square where IR divergencies occur.
However, smoothing the cusps has not only removed the IR divergencies, but
has also removed from the $\tau_i$ integrations
precisely those regions in which the wavefunctions of the asymptotic states 
are completely separated and indeed they are never fully resolved.
In other words the smoothing of the square 
has removed the kinematic poles of the $\tau_i$ integration.
In order to restore them, 
we propose to patch the solution inside the smooth region by matching it to another
solution which has $r$ nearly equal to $r_{IR}=1/Z_{IR} > 0$ and 
extends all the way
to the square boundary. We will see below that this patch is really needed only near the boundary 
(the dotted outer region in Figure 5) and can be 
constructed essentially in flat space.

To restate the argument, we claim that if one were able to 
determine a full solution to the equations of motion in this 
background, the action would separate into a ``bulk'' part described 
by (an IR-regulated) action of ref.\cite{AM}, and a near-boundary contribution which will be constructed shortly.
This near-boundary contribution (in distinction with the AM action) carries dependence on the insertion points $\tau_i$ 
of the vertex operators and is essential for determining the prefactor.
This contribution is manifestly IR-safe (thus can be calculated in a regularisation scheme
of our choice) and due to its $\tau_i$-dependence it is clearly 
distinct from the AM action. 
Finally, this boundary action is not important for the universal exponent of \cite{AM}
(since it will turn out to be formally subleading in $1/\sqrt{\lambda}$), but
is necessary to determine the prefactor.

As already mentioned,
the Alday-Maldacena action
does not depend on the emission points $\tau_i$ and is a homogenous function of external momenta.
Specifically its IR-finite part depends only on the ratios of momenta and as such it does not matter
whether we keep $k$ fixed or $p$ fixed ($k=p/\sqrt{\lambda}$) when we eventually send $\lambda$ to infinity.
There are also 
boundary contributions to the action coming solely from the edges of the square, or more generally
from the edges of $n$-sided polygons, for $n$-point amplitudes. 
Note that in the boundary part of the action it will be
important to keep the {\it physical} unrescaled momenta $p_i^\mu$ fixed when taking the strong coupling limit. 

To calculate the contributions near the edges we
look for the classical solution $X^\mu_{\rm cl}$ near the worldsheet boundary in the dotted
region of Figure 5. There, 
as explained earlier
we treat the $Z$-field as approximately constant $Z=Z_{IR}$.
We want to extremise 
\be
\sqrt{\lambda} S\left[ X,Z \right] 
   + \sum_i  p^\mu_i {X}^\mu \left( \tau_i\right)
    -i  \int d\tau J_\mu \left( 
    \tau\right) \partial_\tau {{X}}^\mu .
    \label{Ltransf}
\ee
For $X^{\mu}$ components, the action \eqref{SZX} is quadratic,
and we can solve for ${X}^{\mu}(\tau,\sigma)$:
\be
2\sqrt{\lambda}\, {X}^{\mu}_{\rm cl} = 
\sum_i G(\tau-\tau_i,\sigma) p_i^\mu +
i\int d\tau' \partial_\tau{G}(\tau-\tau',\sigma) J^{\mu}(\tau')
\label{Xbarmu}
\ee
$G(\tau,\sigma)$ is the Green function of the Laplacian
\be
\frac{\partial}{\partial w} \frac{1}{{Z}^2} \frac{\partial}{\partial \bar{w}}\, G (\tau,\sigma) =
\delta^{(2)} (w) ,
\label{Greenf}
\ee
where $w=\tau+i \sigma$ and $\bar{w}=\tau - i \sigma$. 
We can integrate \eqref{Greenf} and find derivatives of $G$
\be
\frac{\partial}{\partial \bar{w}}\, G (\tau,\sigma) = Z^2 \frac{1}{\bar{w}} \ , \quad
\frac{\partial}{\partial {w}}\, G (\tau,\sigma) = Z^2 \frac{1}{{w}} .
\label{Gder}
\ee
On the boundary, the value of $Z$ is constant $Z_{IR}$, and up to this constant
rescaling, the Green function $G$ in \eqref{Gder} is the same as in flat space.
The effective action \eqref{Ltransf} on our configuration \eqref{Xbarmu}
is given by
\begin{eqnarray}
& & - \frac{1}{2\sqrt{\lambda}} \sum_{ij} p^\mu_i \, G(\tau_i-\tau_j,\sigma=0) \, p_{\mu\,j}\nonumber \\
& & -\frac{i}{2\sqrt{\lambda}} \int d\tau
J_\mu\left( \tau \right)\,  \sum_i \partial_\tau{G} \left( \tau - \tau_i,\sigma=0\right)\,  p_i^\mu
\label{eq:eff} \\
  & &+
\frac{1}{4\sqrt{\lambda}} \int d\tau d\tau^\prime J_\mu\left( \tau \right)\, \partial_\tau^2 {G}\left(
\tau - \tau^\prime,\sigma=0\right)\,  J^\mu\left( \tau^\prime\right) .
\nonumber
\end{eqnarray}
The above expression is evaluated at $\sigma=0$. It is the boundary contribution
 to the action we are after. 
 We denote it 
$\frac{1}{\sqrt{\lambda}} S_{\rm cl}^{\rm boundary}(\tau_i, p_i)$. 
The total action reads
\be
{\sqrt{\lambda}} S_{\rm cl}^{\rm bulk} (p_i)
+ \frac{1}{\sqrt{\lambda}} S_{\rm cl}^{\rm boundary} (\tau_i, p_i) .
\label{Scltot}
\ee
The expression for $\frac{1}{\sqrt{\lambda}} S_{\rm cl}^{\rm boundary}(\tau_i, p_i)$
on the right hand side of \eqref{eq:eff} is precisely the exponent of the generating functional
for the tree-level amplitudes in flat space (we recall that $Z=Z_{IR}=$const and the Green function
is the usual $\log(\tau_i-\tau_j)$). We now take the limit of $\sqrt{\lambda} \to \infty$ keeping
the physical momenta $p_i^\mu$ and the IR-cutoff $Z_{IR}$ fixed. In this limit we pick up the poles
contribution of the tree-level Veneziano amplitude in flat space, which is precisely 
the tree-level Yang-Mills amplitude we are after.
Note that the contribution to the amplitude coming from our patch 
$\frac{1}{\sqrt{\lambda}} S_{\rm cl}^{\rm boundary}(\tau_i, p_i)$ is actually independent of $Z_{IR}$.
This can be seen by recalling that $G= Z_{IR}^2 G_{\rm flat}$ and these two powers of $Z_{IR}$ can
be removed by rescaling $p \to Z_{IR} p$ and $J \to Z_{IR} J$. Since the amplitude $A_n$ goes as
$p^{-n}$, the $Z_{IR}$-dependence disappears from $A_n$. Thus we see that the prefactor is IR-safe
as it should be for the tree-level amplitude. A posteriori, this justifies our use of the 
IR regularisation scheme by a cut-off $Z_{IR}$ for the calculation of the prefactor.

We conclude that
the prefactor $K$ in \eqref{expS} is proportional to the tree-level Yang-Mills amplitude.
It is important to note that the nature of our strong coupling limit is different from the 
one taken in \cite{GM}. The authors of \cite{GM} were taking the combination $\alpha' p_i p_j \to \infty$
which corresponded to their case of interest, namely of taking the very high energy limit of the 
string amplitude in flat space. In our case this procedure would give us the exponential tail
of the Veneziano string amplitude, rather than its pole part. The correct way to take the limit
in the AdS/CFT correspondence case at hand,
is to fix physical momenta and to send $\sqrt{\lambda} \to \infty$ (or $\alpha' p_i p_j \to 0$ in the
language of \cite{GM}).

It is instructive at this point to compare our patched solution approach in $AdS_5$ to a
calculation performed entirely in flat space.\footnote{We are of course not implying that the 
flat space calculation describes the AdS case.
As discussed in detail by Polchinski and Strassler \cite{PS}, the flat
space approximation of the AdS is justified only at high energies, when
the scattering event takes place in a small region of space. Moreover, even
then, the result of the flat space calculation is folded against AdS wave
functions.}
 In the latter case, there is no radial
coordinate ($Z$ or $r$), the action is free
and our ``near-boundary'' solution \eqref{Xbarmu} is valid everywhere in the worldsheet
(with $\sqrt{\lambda}$ replaced by $1/\alpha'$ and $G$ being the flat space Green function, $\log (\tau_i-\tau_j)$).
Thus in flat space we don't need the ``bulk'' solution and the full classical action is given by 
\eqref{eq:eff}
i.e. by the second term in \eqref{Scltot}. The fact that the ``bulk'' action vanishes in flat space 
can be seen explicitly. We have found the configuration $y_0(y_1,y_2)$ which minimises the Nambu-Goto action,
giving $S_{NG}=0$, and satisfies the appropriate boundary conditions.
Hence, 
in flat space the area (or bulk) contribution to the action is trivial, and the
full answer is given by \eqref{eq:eff}. 
This shows an important difference between the classical string actions in the
flat case of \cite{GM} and the AdS case of \cite{AM}. In our language, the Alday-Maldacena AdS action
is the bulk action while the flat-space action is purely of the boundary type.
In the case considered in \cite{GM} the remaining integrations
over emission points $\tau_i$ were performed in the saddle-point approximation relevant to their
case of interest, $\alpha' p_i p_j \to \infty$. As mentioned earlier we work in the 
opposite limit,
$\alpha' p_i p_j \to 0$, where the $\tau_i$ integrations are dominated by regions $\tau_i \to \tau_j$.
This gives the poles part of the Veneziano amplitude
which is precisely the tree-level Yang-Mills amplitude.\footnote{Of course integrations over $\tau_i$
can be carried out exactly giving the full Veneziano string amplitude. Then  the
$\alpha' p_i p_j \to 0$ limit is the Yang-Mills amplitude.}
In the $AdS$ case the bulk action in non-vanishing (and provides for an overall universal exponent),
while the tree-level amplitude gives the
prefactor of the full amplitude. Since this prefactor arises from the ``pinched regions'' where
$\tau_i \simeq \tau_{i+1}$, the flat space calculation is justified. In other words, in this limit the Green function
$G(\tau_i-\tau_j,\sigma=0)$ does not enter the $AdS_5$ bulk and stays on the Minkowski-space boundary.

We now briefly comment on the effect of Gaussian fluctuations 
around the saddle point. A systematic approach to integrate out semiclassical fluctuations 
of strings in $AdS_5\times S^5$ based on the Green-Schwarz formalism \cite{Metsaev:1998it}
was developed in 
Refs.~\cite{Forste:1999qn,Drukker:2000ep}. 
The
bosonic part of the action contains terms quadratic in
\begin{equation}
\left(  \delta^a _b\partial_\alpha + {{\omega_M} ^a}_b \partial_\alpha
{X}_{\rm cl}^M \right) \zeta^b ,
\label{flucts}
\end{equation}
and other similar terms, as explained in \cite{Forste:1999qn,Drukker:2000ep}.
Here $\zeta^a$ denotes quantum fluctuations around the saddle point 
solution ${X}_{\rm cl}^M=(X_{\rm cl}^\mu, Z_{\rm cl}^\mu)$. We note that $J$ and $\tau_i$ can enter
this action only via ${X}_{\rm cl}^\mu$ (i.e. only in the second term in \eqref{flucts}).
Since  $\overline{X}^\mu \sim \lambda^{-1/2} J$, as dictated by \eqref{Xbarmu},
these $J$- and $\tau_i$-dependent terms can be neglected in \eqref{flucts}. 
This line of argument assumes the validity of our patching approach 
and that the scaling of the $J$-dependent part of $\overline{X}^\mu$ is essentially determined by the
flat space patch. 
In this case the overall external-state-dependent structure of the prefactor
at leading order in $\lambda \to \infty$, is not affected by the fluctuations. 
One would need to carry out these
integrations, however, if one wanted to derive $S_0$ in string theory. 
Here we should note that it is not clear if the Alday-Maldacena set-up (at least in dimensional regularisation)
admits a consistent $1/\sqrt{\lambda}$ expansion, as was recently pointed out in Ref.~\cite{Kruczenski:2007cy}.

To summarise, we have argued that the prefactor $K$ of the Alday-Maldacena string amplitude \eqref{expS}
takes essentially the same form as in flat space. Since we are ignoring cubic and higher powers of fluctuations,
and the string worldsheet is a disc, this results in $K$ being proportional to the tree-level planar amplitude
and \eqref{result} follows.

As already mentioned, in gauge theory scattering processes at strong coupling were previously discussed only
for the MHV amplitudes,
i.e. those with $2$ negative and $n-2$ positive helicities. These results follow from the exponentiated ansatz \eqref{BDSans}
of Bern, Dixon and Smirnov \cite{BDS}, which was based on a 3-loop calculation of 4-point amplitudes, 
and was conjectured in \cite{BDS} to hold to all-orders in weakly coupled 
perturbation theory for $n$-point MHV amplitudes.
The prefactor in \eqref{BDSans} is the tree-level Parke-Taylor MHV amplitude \cite{PT,BG}
which for $n$-gluons takes the form 
\be
A_{n ~ \rm MHV}^{~ \rm tree} = g_{YM}^{-2} \frac{\langle p_r p_s \rangle^4}{\langle p_1 p_2 \rangle
\langle p_2 p_3 \rangle \ldots \langle p_n p_1 \rangle} (2\pi)^4 \delta^{(4)} (\sum_i p_i)
\label{PTBG}
\ee
This  amplitude is written in the helicity spinor formalism.
Similar expressions hold for all tree-level MHV amplitudes involving gluons, fermions and scalars of ${\cal N}=4$ SYM. 
They can be found in e.g. section 5 of Ref.~\cite{Khoze}.

$F_n^{\rm BDS}(\lambda;p_i)$ in \eqref{BDSans} 
are functions of kinematical invariants of $n$ external momenta and are computed
perturbatively by Taylor expanding in powers of (small) $\lambda$. The important point is that
the entire kinematic dependence in $F_n^{\rm BDS}(\lambda;p_i)$ is determined from a 1-loop calculation
(i.e. at order $\lambda^1$ in weakly coupled SYM perturbation theory). More precisely, we first 
factor out of the amplitude the IR-divergent part, 
\be
F_n^{\rm BDS}(\lambda;p_i)=F_n^{\rm div}(\frac{1}{\varepsilon};\lambda;p_i) +
F_n^{\rm fin}(\varepsilon;\lambda;p_i) ,
\ee
where the IR-divergent part contains double and single poles
in the $\varepsilon$ parameter of the dimensional reduction. $F_n^{\rm div}$ is fixed by (and is in agreement with)
the general theory of IR divergences in amplitudes \cite{MS,STY,C},\cite{BDS},\cite{AM}. The IR-finite part,
$F_n^{\rm fin}(\varepsilon)$ is determined by the BDS ansatz \cite{BDS} for $\varepsilon=0$
\be
F_n^{\rm fin}(\varepsilon=0;\lambda;p_i)= \frac{f(\lambda)}{4} F_n^{(1)} (p_i) +C(\lambda) .
\label{Finpart}
\ee
If the BDS conjecture is correct, Eq.~\eqref{Finpart} implies that
the kinematic dependence of the
amplitude appears only in $F_n^{(1)} (p_i)$ and is
{\it disentangled} from the coupling $\lambda$ dependence.
The functions
$F_n^{(1)} (p_i)$ are determined at the 1-loop level and are given in Eq.~(4.55) of \cite{BDS}.
Functions $f(\lambda)$ and $C(\lambda)$ depend only on the coupling and are calculated perturbatively \cite{BDS}.
The function $f(\lambda)$ is the soft (cusp) anomalous dimension and is also known in the
strong coupling regime \cite{GKP2,Kruczenski:2002fb,Kruczenski:2007cy}, 
$f(\lambda) \to \frac{\sqrt{\lambda}}{\pi} - \frac{3 \log 2}{ \pi}$. 

In order to have an explicit ansatz for all $n$-point amplitudes we can   
identify the action of the classical string $\frac{\sqrt{\lambda}}{2\pi}{\rm Area}_{cl}$
with the $\sqrt{\lambda}$ terms in the 
$\lambda \to \infty$ limit of the  BDS exponent
for all $n$-point amplitudes. This identification in principle can (and should) be checked explicitly for general $n$
(and in particular for $n \ge 6$) by computing appropriate classical string actions and gauge theory amplitudes as well 
(to check the validity of the BDS ansatz).
 If it does not hold, then even the MHV amplitudes cannot
be matched in string and in gauge theory. 
This would imply that
either the BDS ansatz does not work, or 
that the 
proposal of \cite{AM} of extending the AdS/CFT correspondence to address gauge theory scattering amplitudes 
is incorrect or at least incomplete.
If on the other hand the identification does hold,
we have a formula for all $n$-point planar amplitudes in strongly coupled  ${\cal N}=4$ SYM,
\be
A_n = A_n^{~ \rm tree} e^{F_n^{\rm BDS}(\lambda\to \infty;p_i)} .
\label{string-BDS}
\ee
This would give the prefactor $K$ (at $\varepsilon=0$) for a general amplitude \eqref{expS} in the form,
$K = A_n^{~ \rm tree} e^{- \frac{3 \log 2}{4 \pi}F_n^{(1)} (p_i)}$
up to a numerical coefficient coming from $e^{C(\infty)}$.

After the earlier version of this paper was published
it was pointed out in Refs.~\cite{Drummond:2007cf,Alday:2007he,Drummond:2007bm} that there are reasons to suspect that the BDS conjecture
in gauge theory may fail for MHV amplitudes with $n \ge 6$ external legs. The reason for this lies in the
``dual space'' conformal symmetry which uniquely constrains the $4$-point and $5$-point amplitudes 
to take the BDS form, but not the  $n \ge 6$ point amplitudes which a priori can differ from the BDS proposal
by an arbitrary function of conformal ratios. Since the BDS proposal was influenced by explicit calculations
of only $4$-point and $5$-point amplitudes in \cite{BDS,Cachazo:2006mq,Cachazo:2006tj,Bern:2006vw,Bern:2007ct}  
it may indeed fail or require modifications at the $6$-point level. 
The main conclusions of this paper are not affected by the validity of the specific BDS conjecture.

The statement that in the strong coupling limit all amplitudes of ${\cal N}=4$ SYM exhibit
an exponential form with a universal exponential factor
in our view is the main conclusion 
of the general approach initiated by Alday and Maldacena. The factorised form for these
amplitudes \eqref{result} is our main result.
As mentioned earlier, for non-MHV amplitudes this factorization cannot be seen in the weakly coupled perturbation theory.
However, the remarkable prediction from string theory is that
it must hold in the strong coupling limit $\lambda \to \infty$ where Yang-Mills must simplify dramatically.

\smallskip

\centerline{\bf Acknowledgements} 

We would like to thank Juan Maldacena and Stefan Theisen for valuable comments on the first version of this paper.


\end{document}